\documentclass[twocolumn,showpacs,amsmath,amssymb,prb]{revtex4}

\usepackage{graphicx}
\usepackage{dcolumn}
\usepackage{bm}
\usepackage{psfrag}

\usepackage{color}
\definecolor{red}{rgb}{0.8,0.0,0.0}

\begin{document}

\title{Signature of chirality in scanning-probe imaging of charge flow in graphene}

\author{Matthias Braun}
\author{Luca Chirolli}
\author{Guido Burkard}
\affiliation{Institute of Theoretical Physics C, RWTH Aachen University, D-52056 Aachen, Germany}

\date{\today}

\begin{abstract}
We theoretically propose to directly observe the chiral nature of charge carriers in graphene mono- 
and bilayers within a controlled scattering experiment.  The charge located on a capacitively coupled scanning probe microscope (SPM) tip acts as a scattering center with controllable position on the graphene sheet. Unambiguous features from the chirality of the particles in single and bilayer graphene arise in the ballistic transport in the presence of such a scattering center.   To model the scattering from the smooth potential created by the SPM tip, we derive the space-dependent electron Green function in graphene and solve the scattering problem within the first-order Born approximation.  We calculate the current through a device with a SPM tip between two constrictions (quantum point contacts) as a function of the tip position.
\end{abstract}

\maketitle

The isolation of few and single layer graphene,\cite{Novoselov2004,Novoselov2005,Zhang2005} the two-dimensional carbon allotrope, triggered tremendous research activities (for a review, see Ref.~\onlinecite{Geim2007}). Graphene is technologically of high interest,\cite{Chen2007,Han2007,Lemme2007} as it is chemically stable, but is also very appealing for fundamental scientific research. The electronic eigenstates in graphene obey a linear gapless dispersion relation around the Fermi energy, and bear a pseudospin, which is always aligned parallel in the direction of momentum. This properties imitate the behavior of chiral massles Dirac particles,\cite{Zhang2005,Novoselov2005,Katsnelson2006} discussed in relativistic quantum mechanics. Therefore many concepts of solid-state physics are now being reconsidered for pseudo-relativistic carriers while at the same time, effects known from relativistic quantum mechanics can be found in solid-state physics.  Examples for this are the unusual energies of the Landau levels and the Klein paradox.\cite{Katsnelson2006} Moreover, long spin relaxation lengths \cite{Tombros2007} make graphene an interesting system for spintronics \cite{Hill} and spin-based quantum information processing. \cite{Trauzettel2007}

The high carrier mobility of graphene has lead to an active discussion of impurity scattering. Currently, it is assumed that scattering from Coulomb potentials \cite{DiVincenzo1984, Ando2006, Katsnelson2006b,Biswas2007,Shklovskii2007,Nomura2007,Adam2007,Ostrovsky2006,Peres2006,Katsnelson2007b} limits the conduction electron mobility, while short-ranged defects are less relevant.\cite{Katsnelson2007c}

\begin{figure}[t!]
\includegraphics[width=0.8\columnwidth]{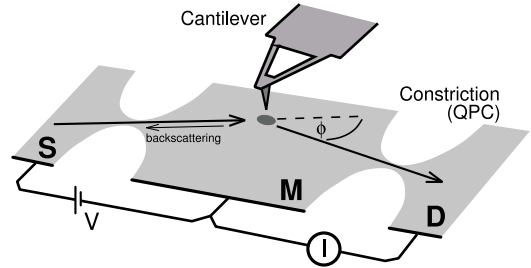}
\caption{\label{fig:scanningsetup}
The proposed setup.  An applied bias voltage $V$ between the source S and the middle region M injects a current into the graphene sheet. The injected electrons scatter from an artificial scatterer created by a SPM tip above the surface. The flux of electrons that are coherently scattered into the drain D can be detected in the drain current I.}
\end{figure}

In the following, we discuss the possibility for a controlled experiment to test whether the charge carriers in graphene behave like chiral particles in a scattering event.
We propose to use the method of mapping electron flow by scanning-probe microscopy (SPM) as
developed and applied to two-dimensional electron gases (2DEGs) in semiconductors
by the Westervelt group.\cite{Topinka2003}
Topinka \textit{et al.} \cite{Topinka2000,Topinka2001} demonstrated that coherent electron flow in a 2DEG formed in a GaAs/GaAlAs heterostructure can be imaged
directly by placing a charged SPM tip on top of the sample.
The tip, being capacitively coupled to the sample, repels the conduction electrons beneath,
forming a circular scatterer with a precisely controllable position. 
By scanning the tip over the conductor, the conductance of the sample is modified depending on the current density beneath.  By setting the SPM tip directly behind a constriction (quantum point contact, QPC),\cite{Topinka2000,Topinka2001} the dominant mechanism of the conductance change is direct backscattering through the QPC next to the source, see Fig.~\ref{fig:scanningsetup}.   In single layer graphene, this backscattering is forbidden,\cite{Katsnelson2006} and therefor one can not expect any resistance change between the respective contacts S and M in such a setting.
To actually use forward scattering from the tip, one might consider putting the tip \textit{in front} of the constriction. Such a setup, however, suffers from the uncontrolled direction of propagation of the incoming particles.
To control the direction of propagation of in and outgoing particles, i.e., the scattering angle, we propose to use two constrictions (QPCs), as in  Ref.~\onlinecite{Aidala2007}.  By applying a voltage $V$ between regions S and M, a current is injected into the middle region (M). The coherent scattering from one QPC to the other gives rise to a measurable current $I$ in the drain D,  which can be precisely measured as a voltage between M and D.\cite{vanHouten1989} The middle region M is supposed to be large, acting as a reservoir, absorbing all electrons that are not scattered into the drain D.  
Such an experiment could directly probe the differential scattering cross section for pseudorelativistic chiral particles and thus it demonstrates the chiral nature of the particles.

For the microscopic description of scattering, and in the closely related Coulomb impurity problem,\cite{Pereira2007} the method of partial-wave expansion was adapted.\cite{Cserti2007,Novikov2007,Katsnelson2007b,Katsnelson2007c} This method is primarily suitable for strong short-ranged potentials such as impurities.
To describe the scattering from the weak potential created by a SPM tip, we apply the method of  first-order Born approximation by deriving the Green functions for single and bilayer graphene, in real space representation.\cite{Novikov2007}

For single layer graphene, the free Dirac Hamiltonian for the envelope wave function at the $K$-point is given by $H=-i\hbar v_F(\sigma_x\partial_x +\sigma_y\partial_y)$ and
in cylindrical coordinates $\bm r = (r,\phi)$ by
\begin{eqnarray}
H= -i \hbar v_F
\left[
\begin{array}{cc}
0 & e^{-i \phi}(\partial_r - \frac{i}{r} \partial_\phi) \\
e^{+i \phi}(\partial_r + \frac{i}{r} \partial_\phi) & 0\\
\end{array}
\right]\,.\label{H1}
\end{eqnarray}
The scattering solution of the Dirac equation
$[H+U(\bm r)]\psi=E_k \psi$, with $E_k=\hbar v_F k$,
can be constructed using the Green function
\begin{eqnarray}\label{eq:greensfunction}
G(\bm \rho)=
-\frac{i}{4} \frac{k^2}{E_k}
\left[
\begin{array}{cc}
H_0(k\rho) & -i H_{-1}(k\rho) e^{-i \theta}\\
i H_{1}(k\rho) e^{i \theta} & H_0(k\rho) \\
\end{array}
\right]\,,
\end{eqnarray}
where $\bm \rho=\bm r-\bm r^\prime=(\rho,\theta)$.
The Green function is a solution of $(H-E)G=-\delta(\bm \rho)$, satisfying
the outgoing radiation condition; thus, the use of the $n$-th order Hankel function
$H_n(z)\equiv H_n^{(1)}$.
The functional form of the Green function is consistent  with Ref.~\onlinecite{Bena2007}.
 Since the Dirac Hamiltonian is formally closely related to the Rashba spin-orbit interaction,
the derivation of the Green function in Eq.~(\ref{eq:greensfunction}) is analogous
to Refs.~\onlinecite{Walls2005,Cserti2006}.

\begin{figure}[t!]
\includegraphics[width=0.8\columnwidth]{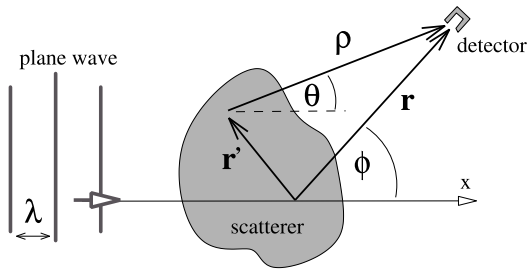}
\caption{\label{fig:exsetup}
Formulation of the scattering problem: An incoming plane wave propagating along the x-direction hits the scattering potential. A detector measures the outgoing flux of the scattered wave as function of the deflection angle $\phi$.}
\end{figure}

The idealized scattering experiment is depicted in Fig.~\ref{fig:exsetup}. A chiral plane wave $\psi_0(\bm r)=e^{i  k  x} [1,1]^T$ propagating along the (so chosen) $x$-axis hits the scattering potential. After the interaction with the potential, a detector measures the flux of the scattered wave as a function of the deflection angle $\phi$.
Within first-order Born approximation, the total wave function $\psi(\bm r)$ of an electron which scatters at a potential $U(\bm r)$ is given by
$\psi(\bm r)= \psi_0(\bm r) +\int d^2r^\prime G(\bm r - \bm r^\prime) U(\bm r^\prime)\psi_0(\bm r^\prime) $. 
Far away from the scattering center the wave function has the asymptotic form
%
$\psi(\bm r)= \psi_0(\bm r)
+f(\phi) \,e^{i k r}\,[1, e^{i \phi}]^T/\sqrt{r}$.
%
The scattered wave describes a chiral circular outgoing wave. The  scattering amplitude is given by
\begin{eqnarray}\label{eq:diffcrosssection}
f(\phi)=-e^{-i\frac{\phi}{2}}\,\sqrt{\frac{i k^3}{2 \pi}}
\cos\frac{\phi}{2}\, \frac{U({\bm q})}{E_k},
\end{eqnarray}
with the Fourier transform $U({\bm q})=\int\!\!d^2r^\prime \,\,U(\bm r^\prime) \,e^{i\bm q\bm r^\prime}$. 
The orientation of the vector $\bm q=k(\bm e_k-\bm e_r)$ of the momentum transfer during the scattering process is determined by the unit vectors $\bm e_r$ and $\bm e_k$ in $r$ and $k$ direction, while its magnitude is given by $q=2 k \sin(\phi/2)$.
The functional form of the scattering amplitude is in close analogy to nonrelativistic two-dimensional scattering.\cite{2D_scattering}
Two main differences appear. First, the  prefactor $e^{-i \phi/2}\cos\frac{\phi}{2}$ leads to the absence of any backscattering at potentials, irrespective of the potential shape.\cite{Novikov2007}  Second, the forward scattering is enhanced by a factor of 2 compared to classical electrons.
In gapped 2DEGs,  scattering at potentials smaller than the Fermi wave length is dominated by s-wave scattering. 
For chiral particles in graphene, the free angular momentum eigenstates bear half-integer angular momentum related to the appearance of a Berry phase of $\pi$.\cite{DiVincenzo1984} Therefore, the scattering amplitude is dominated by states with orbital angular momenta of $+1/2$ and $-1/2$. In backward direction, the interference of the two states is destructive, leading to a suppression of scattering. In forward direction, the interference is constructive, enhancing the scattering amplitude by a factor of 2.
For scatterers larger than the Fermi wavelength (beyond the range of validity of the Born approximation), one can expect that the enhancement of forward scattering becomes non-universal, depending on the details of the potential.

By comparison with the exactly solvable problem of a stepwise constant potential, the range of validity for the Born approximation can be estimated as $(kR)^2 \lesssim E_k/U_0$, where $R$ and $U_0$ denote the characteristic potential size and strength.
As an interesting side note, for the physically relevant Coulomb potential, the first-order Born approximation generates accidentally good results, in three as well as in two dimensions.\cite{2D_scattering}

When we restrict the calculation to circularly symmetric potentials, the scattering amplitude Eq.~(\ref{eq:diffcrosssection}) can be further simplified using
\begin{equation}\label{eq:diffcrosssectionradial}
U({\bm q})=2 \pi\int\!\!dr^\prime \,r^\prime\,J_0(q r^\prime)\, U(r^\prime)\,.
\end{equation}
 The integral in Eq.~(\ref{eq:diffcrosssectionradial}) can be solved analytically for a variety of different scattering potentials, including an unscreened and the exponentially screened Coulomb potential, the stepwise constant potential, potentials of Lorentzian and Gaussian forms, and the potential of a point charge above the graphene sheet.\cite{Gradshteyn2000}

%
%
 

In graphene bilayers, arranged in an A-B (Bernal) stacking, the charge carriers behave like massive gapless Dirac fermions.\cite{McCann2006} To compute the Green function, one can directly make an ansatz motivated by the observation that the operators
$a=-i e^{-i \phi}(\partial_r - \frac{i}{r} \partial_\phi)$ and
$a^\dag=-i e^{i \phi}(\partial_r + \frac{i}{r} \partial_\phi)$ in Eq.~(\ref{H1}) 
act as a type of ladder operators on the Hankel functions, 
i.e., $a^\dag H_n(k\rho) e^{i n \theta}=+i k H_{n+1}(k\rho) e^{i (n+1) \theta}$ and $a H_n(k\rho) e^{i n \theta}=- i k H_{n-1}(k\rho) e^{i (n-1) \theta}$.\cite{note2} 
Note the relative minus sign for $a^\dag$ and $a$.  With this representation, the Hamiltonian for the bilayer as derived in Ref.~\onlinecite{McCann2006} is 
\begin{eqnarray}
H= -\frac{\hbar^2}{2 m}
\left[
\begin{array}{cc}
0 & a a\\
a^\dag a^\dag & 0
\end{array}
\right]\,,
\end{eqnarray}
and the basic functional structure of the Green function for particle energy $E_k=\hbar^2 k^2/2m$ directly follows as
\begin{eqnarray}\label{eq:greensfunction2}
G(\bm \rho)=
-\frac{i}{4}
\frac{k^2}{E_k}
\left[
\begin{array}{cc}
H_0(k\rho) & H_{-2}(k\rho) e^{-2 i \theta}\\
 H_{2}(k\rho) e^{2 i \theta} & H_0(k\rho) \\
\end{array}
\right]\,.
\end{eqnarray}
%
The overall prefactor can be deduced using \cite{Morse1953} that $i/4 (k^2+ \nabla^2) H_0(k\rho)=-\delta(\bm \rho)$ and the fact that $aa^\dag=a^\dag a =-\nabla^2$.
Using the Green function Eq.~(\ref{eq:greensfunction2}), we now construct the scattering wave function.   Within the first-order Born approximation, the wave function takes the asymptotic form $\psi({\bf r})=\psi_0({\bf r})+f(\phi)e^{ikr} \left[1,-e^{2i\phi}\right]^T/\sqrt{r}$. The scattering amplitude $f(\phi)$ is then given by Eq.~(\ref{eq:diffcrosssection}) with the substitution $\phi\,\rightarrow\,2\phi$. 

\begin{figure}[t!]
\includegraphics[width=0.8\columnwidth]{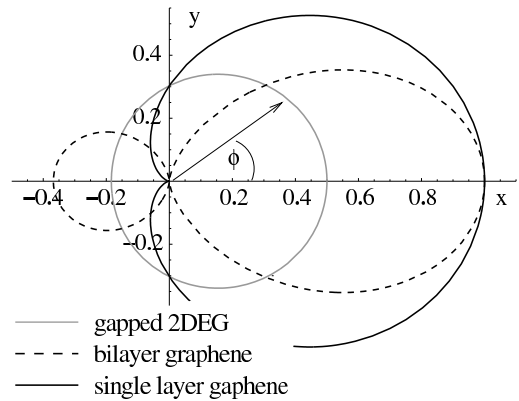}
\caption{\label{fig:crosssection}
Differential cross section normalized by $2 \pi (k R)^4 (U_0/E)^2$ in units of $1/k$ for a circular scatterer of Gaussian shape. 
For single layer graphene, the Berry phase of $\pi$ prohibits backscattering while for bilayer graphene,
it the Berry phase of $2\pi$ prohibits rectangular deflection. 
The radius of the scatterer is chosen $k R=0.5$ for this plot.
}
\end{figure}

%
For a circularly symmetric potential, the scattering amplitude can again be simplified further by using Eq.~(\ref{eq:diffcrosssectionradial}).
The cases of non-relativistic particles in a 2DEG ($j=0$), and of single layer ($j=1$) and bilayer ($j=2$) graphene are distinguished by their respective factors  $e^{-i j \phi/2}\cos(j\phi/2)$ which are due to the Berry phase $j\pi$ acquired during the adiabatic propagation along a closed orbit. For a Gaussian potential $U(\bm r)=U_0 e^{-r^2/2R^2}$ , the resulting cross section $d \sigma/d\phi=|f(\phi)|^2$ becomes
\begin{eqnarray}\label{eq:diffcrosssectionradialb}
\frac{d \sigma}{d\phi}\,\propto \,\,e^{-[2 R k \sin(\phi/2)]^2} \cos^2\frac{j \phi}{2}.
\end{eqnarray}
We plot this result in Fig.~\ref{fig:crosssection}. While the Berry phase prohibits backscattering in single layer graphene, backscattering is allowed in bilayer graphene, while scattering by an angle of $\pm\pi/2$ is forbidden in bilayers.
These results are equivalent to the calculation of the scattering cross section via the k-dependent Green function, see Ref.~\onlinecite{Cheianov2006,Novikov2007}.

\begin{figure}[t]
\includegraphics[width=0.99\columnwidth]{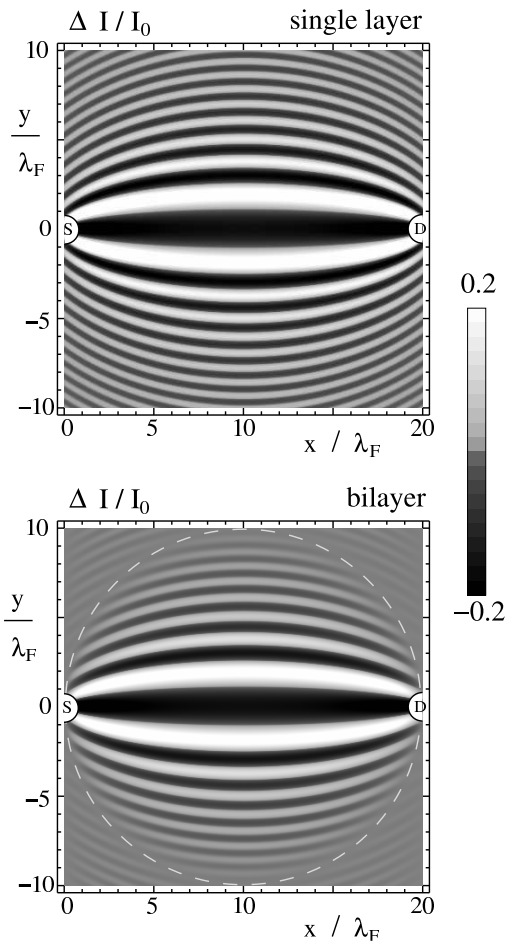}
\caption{\label{fig:currentoplots}
Relative change $\Delta I/I_0$ of the drain (D) current as a function of the tip position $(x,y)$ in units of $\lambda_F$.  The existence of two possible trajectories, one directly from source (S) to drain, the other via the scatterer at $(x,y)$, generates an interference pattern.  Since scattering under an angle of $\pi/2$ is forbidden for a bilayer, a circle without signal (dashed white line) appears. The scattering potential radius and strength were chosen to be $R k=1$, and $U_0/E=0.3$ respectively.
}
\end{figure}

To describe the SPM experiment, as shown in Fig.~\ref{fig:scanningsetup}, we model the potential of the charged tip as a Gaussian. Moreover, we consider the QPC at the source S as point source of chiral electrons (as determined by the Green function) at position $(x,y)=(0,0)$.   In the proximity of the scatterer, we approximate the incoming spherical wave $\psi$ (centered around the QPC) as a plane wave and derive the scattered outgoing spherical wave (centered around the SPM-tip). The drain current $I$ is calculated as the current at the location of the drain QPC at  $(x,y)=(d,0)$, where $d$ is the distance between the source and drain QPCs.  The normal component of the current is given by $J_x=v_F\psi^{\dag}\sigma_x\psi$ for single layer, and by $J_x=-\frac{\hbar}{m} {\rm Im} [\psi^{\dag}(\sigma_x\partial_x+\sigma_y\partial_y)\psi]$ for bilayer graphene. Here $\psi$ labels the sum of the incoming and outgoing spherical waves. In Fig.~\ref{fig:currentoplots}, we plot the relative  change of the drain current $\Delta I/I_0$ due to the presence of the tip as a function of the tip position $(x,y)$.
Here, $\Delta I=I-I_0$ where $I_0$ is the current in the absence of the SPM tip.

In the presence of the scatterer, two ballistic trajectories lead from source to drain: either the electrons travel directly form the source (S) to the drain (D) or they scatter at the SPM tip potential and from there into the drain (D). The spatial pattern due to the interference between these two trajectories reveals the Fermi electron wavelength $\lambda_F=2\pi/k=hv_F/E_F$, the degree of coherence,  as well as the scattering phase.\cite{Leroy2005}

The required scattering angle to pass from source to drain is a function of the tip position relative to source and drain. Therefore this experiment realizes to some extent an angle-resolved measurement of  $d\sigma/d\phi$. We assume the middle region (M) of the graphene sample to be large (but not larger than the coherence length $l_\phi$), i.e., $R\ll d\ll l_\phi$, so that scattering events with other angles will not significantly contribute to the drain current. 
For single layer graphene, one can expect a rapid loss of signal if $\phi>\pi/2$, as scattering by larger angles is strongly suppressed. For bilayer graphene, one can expect forward as well as backscattering. However, a scattering angle of $\phi=\pi/2$ is forbidden. Therefore, a circular line appears (according to Thales' theorem), 
indicating a total absence of scattering.
In contrast, for a conventional semiconductor-based 2DEG, the intensity distribution is much more homogeneous.

As shown in recent experiments,\cite{Gildemeister2007} the minimal 
tip-induced potential width is about 300 nm at a potential height of 1 meV. 
At a Fermi energy of 10 meV in the graphene sheet, the range of validity 
of the first-order Born approximation $(kR)^2 \lesssim E_k/U_0$ is violated 
by about one order of mangitude. This violation leads to deviations mainly for 
forward scattering. Since in the proposed experiment the large angle scattering is of primary 
interest, first-order Born approximation is still expected to deliver qualitatively 
correct results.

Furthermore, sample roughness and disorder will also significantly modify this idealized experimental result, as already observed in 2DEGs.\cite{Topinka2000,Topinka2001,Topinka2003} Even so, if one assumes that small-angle disorder scattering is dominant, then disorder will be of most importance, when the tip is directly in between the source and the drain. With increasing scattering angle, the experimental result will be more and more robust against weak disorder. 

Our calculation was done for one of the two degenerate valleys (the Dirac point at momentum $K$).  For
the other valley ($K^\prime$), only the sign of $\phi$ must be reversed in Eq.~(\ref{H1}).
Therefore, the results for the scattering cross section Eq.~(\ref{eq:diffcrosssectionradialb})
and the current (Fig.~\ref{fig:currentoplots}) remain unchanged for $K^\prime$, as they are even functions
of $\phi$.  Therefore, we expect these results to persist for arbitrary incoherent mixtures
of $K$ and $K^\prime$ without any loss of interference visibility.

\begin{figure}[t!]
\includegraphics[width=0.8\columnwidth]{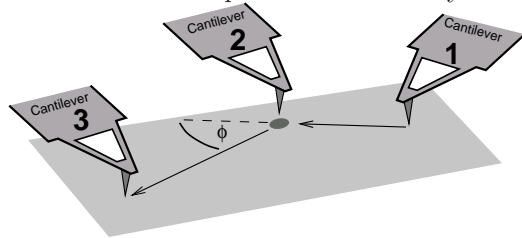}
\caption{\label{fig:3tips}
Alternative setup where the QPCs have been replaced by two additional SPM tips playing the role of the source and drain contacts.}
\end{figure}

To test the ballistic current through a graphene sheet, it would also be possible to use a multi-tip setup, see Fig.~\ref{fig:3tips}, as developed recently.\cite{Jaschinsky2006}
Thereby, the roles of the source and drain contacts are played by two additional SPM tips.
The SPM tip which is capacitively coupled and creates the scattering potential can also be moved behind the other tips and thus allows for a mapping of the complete angular dependence of the scattering amplitude.  The small cross sections of the contacts on the graphene ensure the angular resolution of the scattering experiment. This alternative setup, however, bears the experimental disadvantage that the resistance change as function of the scatterer position can not be measured in a non-local geometry as in the case of the QPC-setup. 

In conclusion, we propose to test the chirality of electrons in graphene mono- and bilayers in potential scattering, probed by a SPM tip in a transport setting with two QPCs. We describe the scattering within first-order Born approximation, which requires the derivation of the electron Green function in real space.

We thank R. M. Westervelt and M. Morgenstern for discussions. This work was supported in part by the Swiss SNF.


\begin{thebibliography}{99}
\bibitem{note2}
Note that  $e^{-i \phi}(\partial_r - \frac{i}{r} \partial_\phi)H_n(k\rho) e^{i n \phi} =e^{-i \theta}(\partial_\rho - \frac{i}{\rho} \partial_\theta)H_n(k\rho) e^{i n \phi}$ due to translational invariance.
\end{thebibliography}
\end{document}